\documentclass[runningheads,openany]{svmult}

\usepackage{makeidx}   
\usepackage{graphicx}  
\usepackage{subeqnar}  
\usepackage{multicol}  
\usepackage{physprbb}  
\makeindex             

\usepackage{epsfig}
\usepackage{latexsym}

\newcommand{\klesssim}{\mathrel{\hbox{\rlap{\hbox{\lower4pt\hbox{$\sim$}}}\hbox{$<$}}}}

\newcommand{\llsim}{\mathrel{\lower4pt\hbox{$\sim$}}\hskip-11.5pt\raise1.6pt\hbox{$<$}\;}

\begin{document}

\title*{Precision Spectroscopy of  Atomic Hydrogen  \protect\newline
and Variations of Fundamental Constants
 }
\toctitle{Precision Spectroscopy of  Atomic Hydrogen and
Variations of Fundamental Constants}

\titlerunning{Precision Spectroscopy of Atomic Hydrogen}

\titlerunning{Precision Spectroscopy of Atomic Hydrogen}

\author{M.~Fischer\inst{1}
\and N.~Kolachevsky \inst{1,2} \and M.~Zimmermann \inst{1} \and
R.~Holzwarth \inst{1} \and Th.~Udem \inst{1}
  \and T.W.~H\"{a}nsch \inst{1,3}
  \and M.~Abgrall \inst{4}
  \and J.~Gr\"{u}nert \inst{4}
  \and I.~Maksimovic \inst{4}
\and S.~Bize \inst{4} \and H.~Marion \inst{4} \and F.~Pereira Dos
Santos \inst{4} \and P.~Lemonde \inst{4} \and G.~Santarelli
\inst{4} \and P.~Laurent \inst{4}  \and A.~Clairon \inst{4} \and
C.~Salomon \inst{5}}
\authorrunning{M.~Fischer et al.}

\institute{ Max-Planck-Institut f\"{u}r Quantenoptik,
Hans-Kopfermann-Stra{\ss}e 1, 85748 Garching, Germany \and  P.N.
Lebedev Physics Institute, Moscow, Russia \and
Ludwig-Maximilians-University, Munich, Germany \and BNM-SYRTE,
Observatoire de Paris, 61 Avenue de l'Observatoire, 75014 Paris,
France \and Laboratoire Kastler Brossel, ENS, 24 rue Lhomond,
75005 Paris, France}

\maketitle

\begin{abstract}

In 2003 we have  measured the absolute frequency of the $(1S, F=1,
m_F=\pm 1)\rightarrow (2S, F'=1, m_F'=\pm 1)$ two-photon
transition in atomic hydrogen. By comparison with the earlier
measurement in 1999 we can set an upper limit on its variation of
$(-29\pm 57)$~Hz within 44 months. We have combined this result
with recently published results of optical transition frequency
measurements in the $^{199}$Hg$^+$ ion and comparison between
clocks based on $^{87}$Rb and $^{133}$Cs. From this combination we
deduce the limits for fractional time variations of the fine
structure constant $\dot{\alpha}/\alpha=\partial/{\partial t}(\ln
\alpha)=(-0.9\pm 2.9)\times 10^{-15}$~yr$^{-1}$ and for the ratio
of $^{87}$Rb and $^{133}$Cs nuclear magnetic moments
$\partial/{\partial t}(\ln[\mu_{\rm {Rb}}/\mu_{\rm
{Cs}}])=(-0.5\pm 1.7)\times 10^{-15}$~yr$^{-1}$. This is the first
precise restriction for the fractional time variation of $\alpha$
made without assumptions about the relative drifts of the
constants of electromagnetic, strong and weak interactions.
\end{abstract}

\section{Introduction}

The question of constancy of fundamental constants was first
raised in Dirac's ``Large Number hypothesis'' (1937)  which aimed
for a harmonization of basic laws of physics \cite{FischDir37}.
Since then, this hypothesis has been reviewed and extended by many
other scientists opening a broad field of theoretical and
experimental investigations. As there is no accepted theory
predicting the values of fundamental constants, the question of
their possible time variation belongs mostly to the field of
experimental physics. The last decades saw a number of different
astrophysical, geological, and laboratory tests searching for
their possible variation in different time epochs with an ever
increasing accuracy. From the point of view of its importance for
physics in general, this problem stays at the same level as the
test of  $CPT$-symmetry and the search for an electric dipole
moment of elementary particles.

In all metric theories of gravity including general relativity any
drift of non-gravitational constants is forbidden. This statement
bases on Einstein's Equivalence Principle (EEP) postulating that
(i) the weight of a body is proportional to its mass, (ii) the
result of any non-gravitational measurement is independent of the
velocity of the laboratory rest-frame (local Lorentz invariance),
and (iii)~the result of a non-gravitational measurement is
independent of its time and position in this frame (local time and
position invariance). On the other hand, theories towards a
unified description of quantum mechanics and gravity allow for, or
even predict some violations of EEP \cite{FischDamour}. In this
sense, any experimental search for a drift of fundamental
constants tests the validity of EEP as well as it provides
important constraints on new theoretical models.

\index{Fundamental~constants!correlations}

The basic principle of all tests of the stability of fundamental
constants is the investigation of time variations of some stable
physical value $\Theta$. Usually, $\Theta$~is a dimensionless
value which can be the ratio of reaction cross-sections, the
distances, masses, magnetic moments, frequencies and so on. In an
experiment one measures the value $\Theta$ at two different times
$t_1$ and $t_2$ and compares $\Theta(t_1)$ with $\Theta(t_2)$. The
value of $\Theta$ may depend on a number of fundamental constants
$\alpha_i$ ($i=1,\ldots ,n$) and the conclusion about drifts of
$\alpha_i$ originate from the analysis of
$\Theta(t_1)-\Theta(t_2)$. The functional connection between
$\Theta$ and $\alpha_i$ can include rather complicated theoretical
models and assumptions which make the results somehow unclear and
strongly model-dependent. Even if the dependence
$\Theta(\alpha_i)$ is straightforward, it is difficult to separate
the contributions from individual $\alpha_i$ drifts if $n>1$. As
mentioned in Ref. \cite{FischKarsh}, all the relative drifts of
fundamental constants, if existing, should be on the same order of
magnitude which can result in a cancelation of the drift of
$\Theta$ as well as in its amplification. For example, according
to an elaborate scenario in the framework of a Grand Unification
Theory, the fractional time variation of hadron masses and their
magnetic moments should change about 38 times faster than the
fractional time variation of the fine structure constant $\alpha$
\cite{FischCal02}.

Astrophysical and geological methods test the stability of
fundamental constants over very long time intervals of 1--10 Gyr.
Due to the large difference of $|t_1-t_2|$, the sensitivity of
these methods to a monotonic long-time drift is very high but they
are insensitive to more rapid fluctuations. A recent analysis of
quasar absorption spectra by Murphy {\it et al.} with redshifted
UV transition lines indicates  a variation of $\alpha$ on the
level of $\Delta \alpha/\alpha = (-0.54 \pm 0.12)\times 10^{-5} $
in the first half of the evolution of the universe (5--11 Gyr ago)
\cite{FischMur03}. There are also indications that in this period
the electron to proton mass ratio was different from its
contemporary value on the same level of $10^{-5}$
\cite{FischVar02}. The analysis of astrophysical data requires a
number of model assumptions which include not only the
well-established scenarios of the  evolution of the universe, but
also assumptions about the isotopic abundance in interstellar gas
clouds, the presence of magnetic fields and others (see e.g. the
review~\cite{FischUzan}) which are difficult to prove. More recent
observation of quasar absorption spectra, performed by different
groups, seem to rule out a variation of $\alpha$ on the level
observed by Murphy {\it et al.} \cite{FischChand,FischQuast}.

A very stringent limit for the time variation of $\alpha$ on
geological timescales follows from the analysis of isotope
abundance ratios in the natural fission reactor of Oklo, Gabon,
which operated about $2$~Gyr ago. A recent re-analysis of the data
of the $^{149}$Sm/$^{147}$Sm isotope abundance ratio sets a limit
of $\Delta \alpha/\alpha = (-0.36 \pm 1.44)\times
10^{-8}$~\cite{FischFuj00}. The interpretation of the data is not
unambiguous, as the result strongly depends on reactor operating
conditions which are not exactly known. Selecting another possible
reaction branch yields a value of $\Delta \alpha/\alpha = (9.8 \pm
0.8)\times 10^{-8}$~\cite{FischFuj00}. In contrast to the first
one, this result indicates a non-zero drift.

Laboratory experiments are sensitive to variations of fundamental
constants during the last few years and typically base  on precise
frequency measurements in atomic or molecular systems. In
comparison to astrophysical and geological ones, laboratory
measurements considerably win in relative accuracy which, in spite
of much shorter $|t_1-t_2|$ time intervals, leads to a competitive
sensitivity on drifts. Moreover, in this case systematic effects
can be well controlled and the  dependence of the transition
frequencies on fundamental constants is straightforward.

 Any
absolute frequency measurement of some transition in an atomic
system is a comparison of this frequency with the frequency of the
ground state hyperfine transition of $^{133}$Cs. Such a
measurement of~{\emph { one}} transition frequency in~{\emph {
one}} atomic system imposes a limit on the variation of some
simple combinations of $\alpha$, nuclear and electron magnetic
moments and/or their masses~\cite{FischKarsh}. To separate the
drift of the fine-structure constant one needs either to impose
some restricting assumptions on the time dependence of the
coupling constants of the strong ($\alpha_S$) and electroweak
($\alpha_W$) interaction~\cite{FischCal02,FischPrest95} or make
absolute measurements of {\emph {two or more}} transition
frequencies possessing {\emph {different functional dependencies}}
on the fundamental constants. The second method does not include
any model parameters or additional assumptions which favorably
distinguishes it. It is also possible to make such a
model-independent evaluation by directly comparing e.g. gross- and
fine structure or two gross-structure frequencies without
comparison to a primary frequency standard and thus excluding the
corresponding dependence on the nuclear magnetic moments. To our
knowledge, such laboratory experiments still have not been done
with a level of accuracy competing modern  absolute frequency
measurements.

In this work we deduce separate stringent limits for the relative
drifts of the fine structure constant $\alpha$ and the ratio
$\mu_{\rm {Cs}}/\mu_B$ by combining the results of two optical
frequency measurements in the hydrogen atom and in the mercury ion
 relative to the ground state hyperfine splitting of
$^{133}$Cs. The measurements of the hydrogen transition frequency
have been carried out at MPQ, Garching, Germany and are described
below. The experiments on the drift of the $5d^{10}6s \ ^2S_{1/2}
(F=0) \rightarrow 5d^9 6s^2\ ^2D_{5/2} (F'=2, m'_F=0)$ electric
quadrupole transition frequency $\nu_{\rm {Hg}}$ in $^{199}$Hg$^+$
have been performed by the group of J.~Bergquist at NIST, Boulder
CO, USA between July 2000 and December 2002. They are described in
detail elsewhere \cite{FischBiz03}.

 From 1999 to 2003, the
ratio of the ground state hyperfine splittings of $^{87}$Rb and
$^{133}$Cs has been determined from a comparison between several
simultaneously running atomic fountain clocks in BNM-SYRTE and
ENS, Paris, France \cite{FischMar03}. Using this result, we can
also set a limit for the fractional time variation of the Rb and
Cs nuclear magnetic moment ratio $\mu_{\rm {Rb}}/\mu_{\rm {Cs}}$.

As the measurements were performed at different places and at
different times we have to use the hypothesis, that the results
are independent of the place on the Earth's orbit, at least within
the last 4 years. In other words, we have to assume a validity of
 local Lorentz invariance (LLI) and local position invariance
(LPI) as well as to make the additional hypothesis, that the
constants change on a cosmological time scale and do not oscillate
within a few years (linear drifts). With this exceptions, our
results are independent of any further model assumptions like any
form of correlation between the constants or constancy of a
particular set of constants.

\begin{figure}[b!]
\begin{center}
\includegraphics[width=0.95\textwidth]{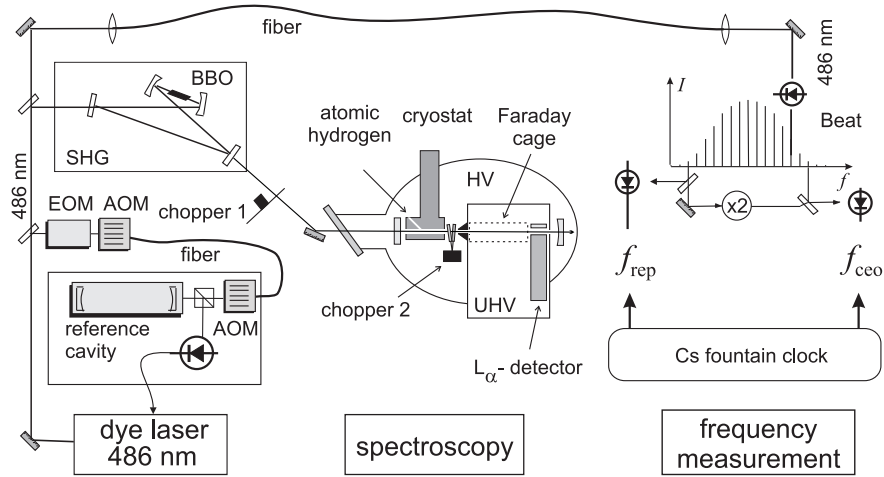}
\caption{ Experimental setup for the  comparison of the hydrogen
$1S$--$2S$ transition frequency with a primary frequency standard.
 The 486~nm light is doubled in a Barium
$\beta$-Borate crystal (BBO) in the second harmonic generation
(SHG) stage. Resulting radiation is coupled to a linear
enhancement cavity in a vacuum chamber with the pressure of about
$10^{-5}$~mbar (HV), while the excitation and detection take place
in an ultra-high vacuum (UHV) zone at the pressure of
$10^{-8}$--$10^{-7}$~mbar. EOM and AOM denote electro- and
acousto-optical modulators correspondingly. } \label{fischer1}
\end{center}
\end{figure}

\section{Hydrogen spectrometer}

\index{Two-photon transition in hydrogen} \index{Doppler-free
spectroscopy} \index{Atomic beam, laser spectroscopy}

 In 1999 \cite{FischNie00a} and 2003, the
frequency of the $(1S, F=1, m_F=\pm 1) \rightarrow (2S, F'=1,
m_F'=\pm 1)$ two-photon transition in atomic hydrogen has been
phase coherently compared to the frequency of the ground state
hyperfine splitting in $^{133}$Cs using a high-resolution hydrogen
spectrometer and a frequency comb technique \cite{FischRei00a}. In
1999, the accuracy of the evaluation of the transition frequency
was $1.8\times10^{-14}$. The setup of the hydrogen spectrometer
used during this measurement has been described previously in
\cite{FischHub98b}. We have introduced a number of improvements in
the spectroscopic setup which will be described in the following.
A sketch of the actual setup is shown in Fig.\ref{fischer1}.

\begin{figure}[t]
\begin{center}
\includegraphics[width=0.95\textwidth]{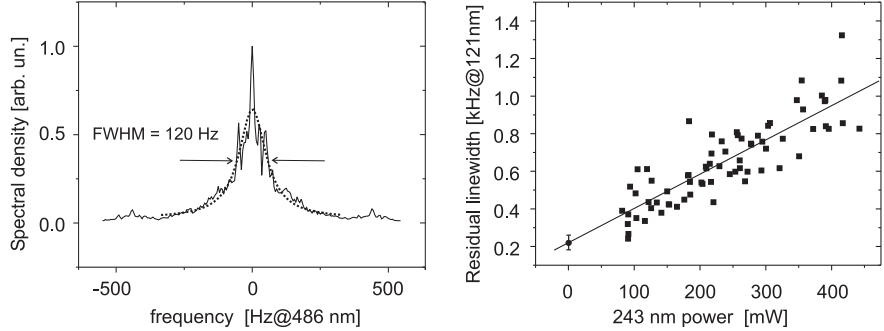}
\caption{  (left) spectrum of the beat signal between laser fields
locked to two independent cavities. (right) extrapolation of the
ionization broadening of the $1S$--$2S$ transition spectra to zero
excitation power circulating in the enhancement cavity. }
\label{fischer2}
\end{center}
\end{figure}

A cw dye laser emitting near 486 nm is locked to an external
reference cavity. The cavity used during the 1999 measurement was
made from Zerodur and had a typical drift of 25~Hz~s$^{-1}$ at the
fundamental frequency. The new cavity made from Ultra Low
Expansion (ULE) glass for the 2003 measurement is better shielded
against the environment. Its drift has been less than
0.5~Hz~s$^{-1}$ for the entire time of the measurement. Due to the
better thermal and acoustic isolation and improvements in the
laser locking electronics, the laser linewidth is narrower than it
has been in 1999. An upper limit for the laser linewidth has been
deduced from an investigation of the beat signal between two laser
fields locked separately to independent Zerodur and ULE cavities.
The spectrum of 12 averaged scans, each taken in 0.2~s is
represented in Fig.\ref{fischer2} (left). The width of this beat
signal spectrum is about  120~Hz at a laser wavelength of 486~nm.
Yet it is impossible to distinguish between the individual noise
contributions from the two independent, but not equivalent
cavities. Another restriction can be deduced from the analysis of
the $1S$--$2S$ transition spectra. The linewidth of the transition
is mainly defined by time-of-flight broadening and is between
1~kHz and 5~kHz at 121~nm. The residual linewidth obtained after
subtracting the estimated contribution of time-of flight
broadening is plotted on Fig.\ref{fischer2} (right) versus the
excitation light power. The observed broadening is due to the
ionization processes and corresponds to a reduced lifetime of the
metastable excited atoms. Extrapolating the residual linewidth to
zero intensity, we get 240(30)~Hz at 121~nm. This can be
considered as a contribution from laser frequency fluctuations.
Thus, we evaluate the 486~nm laser linewidth as 60 Hz for
averaging times of 0.5~s.

A small part of the laser light is transferred to the neighboring
laboratory via a single mode fiber where its absolute frequency
can be measured. The main part is frequency doubled in a BBO
crystal. For higher conversion efficiency, the crystal is placed
in a folded enhancement cavity. The resulting 20 mW of radiation
near 243 nm (corresponding to half of the $1S$--$2S$ transition
frequency) is coupled into a linear enhancement cavity inside the
vacuum chamber of the hydrogen spectrometer.

Molecular hydrogen is dissociated in a 15~W, 2.5~GHz
radio-frequency gas discharge. The resulting flow of atomic
hydrogen is cooled by inelastic collisions with the walls of a
copper nozzle having the temperature of 5--7~K. The nozzle forms a
beam of cold atomic hydrogen which leaves the nozzle collinearly
with the cavity axis and enters the interaction region between the
nozzle and the $L_\alpha$-detector. This region is shielded from
stray electric fields by a Faraday cage. Some of the atoms are
 excited from the ground state to the metastable $2S$
state by Doppler-free absorption of two counter-propagating
photons from the laser field in the enhancement cavity. After the
1999 measurement which had been performed at a background gas
pressure of around $10^{-6}$ mbar in the interaction region, we
have upgraded the vacuum system to a differential pumping
configuration. This allows us to vary the background gas pressure
between  $10^{-8}$ and $10^{-7}$ mbar in 2003 and to reduce the
background gas pressure shift and the corresponding uncertainty
down to 2~Hz.
\begin{figure} [t!]
\begin{minipage}[b]{0.5\textwidth}
\begin{center}
\includegraphics[width=0.95\textwidth]{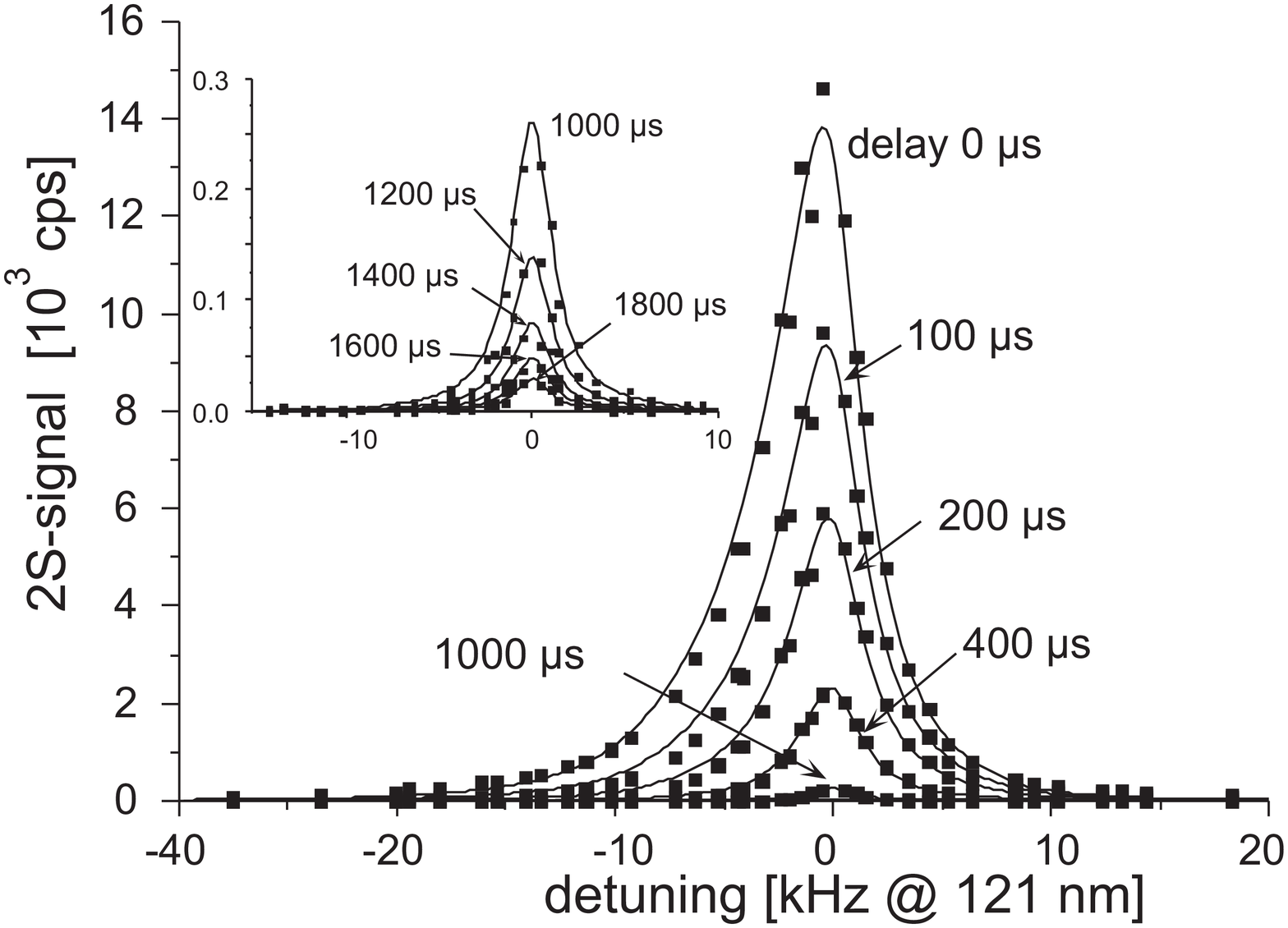}
\end{center}
\end{minipage}\hfill
\begin{minipage}[b]{0.5\textwidth}
\begin{center}
\includegraphics[width=0.95\textwidth]{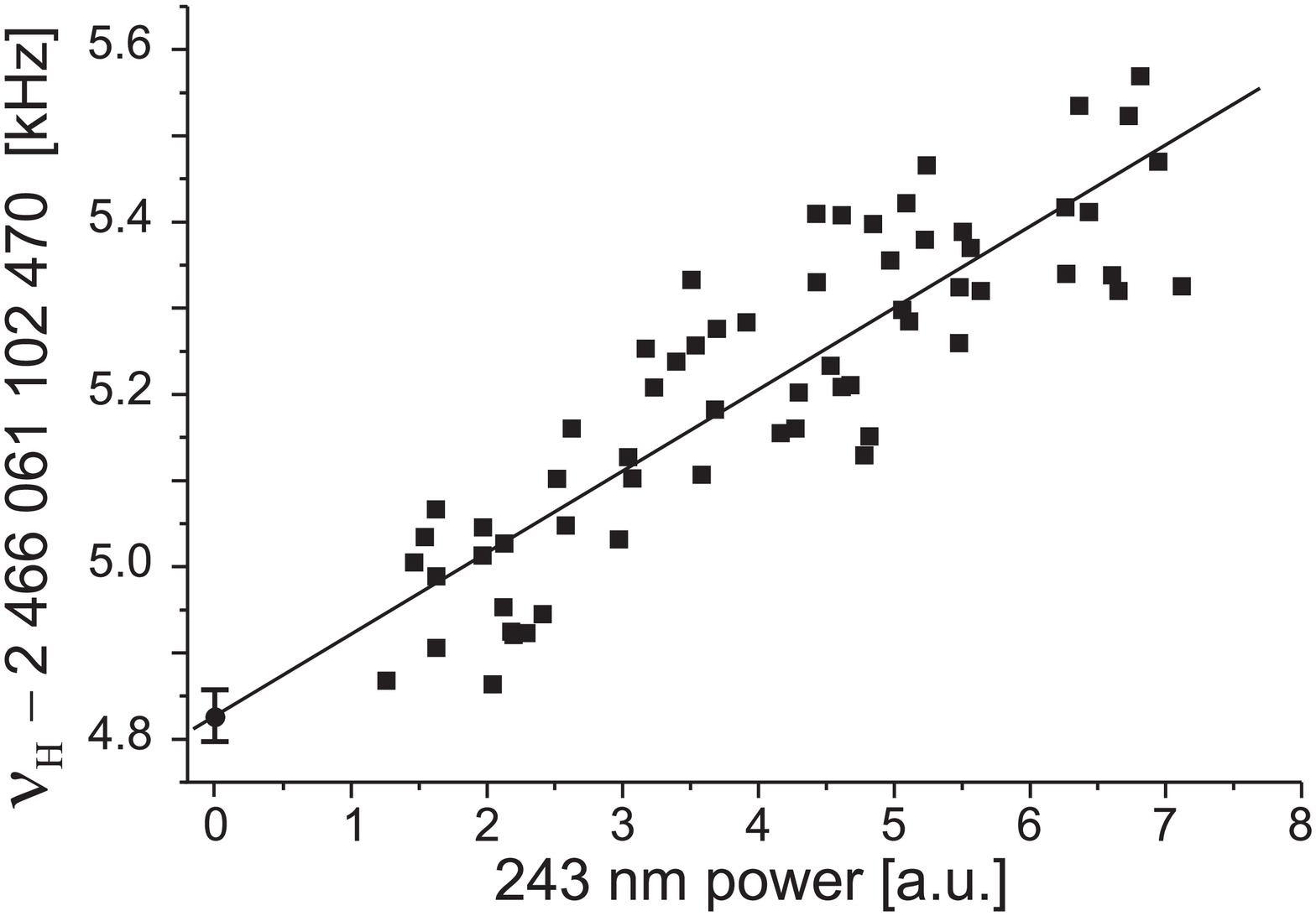}
\end{center}
\end{minipage}\hfill
\caption{ (left) simultaneous fit of a $1S$--$2S$ transition
spectrum recorded at different delays $\Delta t$. The nozzle
temperature was equal to 7~K. (right) AC Stark shift
extrapolation.}\label{fischer3}
\end{figure}

Due to small apertures, only atoms flying close to the cavity axis
can enter the detection region where the $2S$ atoms are quenched
in a small electric field and emit $L_{\alpha}$-photons. The
excitation light and the hydrogen beam are periodically blocked by
two phase locked choppers operating at 160~Hz frequency and the
$L_{\alpha}$-photons are counted time-resolved only in the dark
period of the cycle. This eliminates background counts from the
excitation light. The delay $\Delta t$ between blocking the 243 nm
radiation and the start of counting sets an upper limit on the
velocity of the atoms which contribute to the signal. For some
definite $\Delta t$ only atoms with velocities $v < d/\Delta t$
are selected, where $d$ is the distance between nozzle and
detector. Therefore, velocity dependent systematic effects such as
the second-order Doppler shift and the time-of-flight broadening
are smaller for spectra recorded at larger $\Delta t$. The
hydrogen beam is blocked by a fork chopper in less then 200~$\mu$s
after the blocking of the excitation light to  prevent slow atoms
from being blown away by fast atoms that emerge subsequently from
the nozzle. With the help of a multi-channel scaler, we count all
photons and sort them into 12 equidistant time bins. From each
scan of the laser frequency over the hydrogen $1S$--$2S$ resonance
we therefore get 12 spectra at different delays. To correct for
the second order Doppler shift, we use an elaborated theoretical
model \cite{FischHub98b} to fit all the delayed spectra of one
scan simultaneously with one set of 7 fit parameters (see
Fig.\ref{fischer3}). The result of the fitting procedure is the
$1S$--$2S$ transition frequency for the hydrogen atom at rest.

Besides the second order Doppler effect, the other dominating
systematic effect is the dynamic AC Stark shift which shifts the
transition frequency linearly with the excitation light intensity.
We have varied the intensity and extrapolate the transition
frequency to zero intensity to correct for it \cite{FischNie00a}.
A typical set of data taken within one day of measurement in 2003
and the corresponding extrapolation is presented in
Fig.\ref{fischer3} (right).

\section{Frequency measurement}

\index{Precision frequency measurements!optical}
\index{Frequency~comb!octave spanning}

For an absolute measurement of the $1S$--$2S$ transition frequency
in units of Hz, the frequency of the dye laser near 616.5~THz
(486~nm) was phase coherently compared with a cesium fountain
clock \cite{FischMar03}. To bridge the large gap between the
optical- and radio-frequency (RF) domain we took advantage of the
recently developed femtosecond laser frequency comb technique
incorporating a highly nonlinear glass fiber, which allows for a
further simplification of the experimental setup as compared to
the measurement performed in 1999. In this section we give an
introduction of the frequency comb technique and a description of
the experimental setup, which was used for $1S$--$2S$ frequency
measurement in 2003.

The pulse train emitted by a sufficiently stable mode locked
femtosecond (fs) laser equals a comb of cw laser modes in the
frequency domain. The frequency of each mode of this comb can be
written as $f_n=n f_{\rm rep}+f_{\rm ceo}$, where $f_{\rm rep}$ is
the pulse repetition rate of the fs laser, $n$ is an integer
number and $f_{\rm ceo}$ is the so-called carrier envelope offset
frequency \cite{FischRei99}.
\begin{figure}[t!]
\begin{center}
\includegraphics[width=0.95\textwidth]{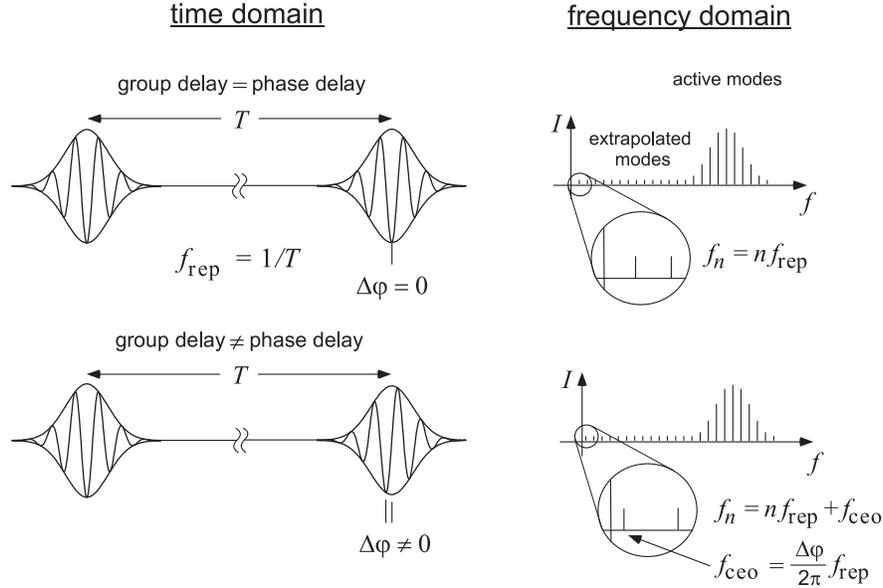}
\caption{ Time- and frequency domain representation of a pulse
train emitted by a mode-locked laser. If the phase delay is
different from the group delay inside the laser cavity, this leads
to the so-called carrier envelope offset frequency $f_{\rm ceo}$,
which shifts the frequency comb as a whole. } \label{fischer5}
\end{center}
\end{figure}

\index{Frequency~comb!repetition rate frequency}
 \index{Frequency~comb!carrier envelope frequency}

 The fs laser emits a train
of pulses with a repetition rate $f_{\rm rep}=1/T$, where $T$ is
the time between consecutive pulses. The envelope function of the
pulses has the periodicity of $f_{\rm rep}$, but it does not
necessarily mean that the electrical field of the pulses has the
same periodicity.  The pulses have
 identical field transients only when the laser cavity roundtrip  phase delay of
the fs laser pulse equals the group delay (Fig.\ref{fischer5}
top). In this case not only the envelope function but also the
electrical field has the periodicity of $f_{\rm rep}$. This leads
to a Fourier spectrum $f_n=n f_{\rm rep}$, where all the modes are
exact multiples of $f_{\rm rep}$. Generally, the group delay does
not equal the phase delay inside the cavity and the frequencies
$f_n$ cannot be integer multiples of $f_{\rm rep}$
(Fig.\ref{fischer5} bottom). Denoting the phase shift between the
envelope function and the carrier frequency of consecutive pulses
as $\Delta\varphi$ one can show, that the frequencies can be
written as
\begin{eqnarray}
f_n = nf_{\rm rep} + f_{\rm ceo}\qquad \textrm{with}\\ \nonumber
 f_{\rm ceo} =\frac{\Delta\varphi}{ 2\pi}\, f_{\rm rep}\, , \qquad  f_{\rm ceo} < f_{\rm rep}\, .
\end{eqnarray}

If $f_{\rm rep}$ and $f_{\rm ceo}$ are fixed, all the modes of the
frequency comb are determined in their frequency and can be used
for measuring the frequency of cw laser light via beat notes
between the cw laser light and a nearby comb mode. The large gap
between the RF and the optical domain is bridged due to the fact
that $n$ is a large integer number of the order of $10^6$. To use
the frequency comb for high precision optical frequency
measurements one has to link $f_{\rm rep}$ and $f_{\rm ceo}$ phase
coherently to a Cs primary frequency standard. The Cs clock
provides us with an extremely precise reference frequency to
control $f_{\rm rep}$ and $f_{\rm ceo}$. The pulse repetition rate
$f_{\rm rep}$ is easily measured with a photodiode and controlled
via the length of the fs laser cavity, which can be changed by
means of a piezo-mounted  cavity mirror. In general, $f_{\rm ceo}$
can be controlled by adjusting the pump power of the fs laser
\cite{FischHau01,FischHol00}. In the case of a linear laser cavity
with a prism pair to compensate for the group velocity dispersion,
$f_{\rm ceo}$ can also be {\em controlled} by tilting the end
mirror of the dispersive arm of the laser cavity
\cite{FischRei99}. The challenging problem for some time was to
 {\em measure}
$f_{\rm ceo}$. If the spectrum of the optical frequency comb
covers an entire octave, $f_{\rm ceo}$ is most conveniently
determined by frequency doubling the mode $f_n$ on the low
frequency side of the comb spectrum and comparing the result with
the mode $f_{2n}$ on the high frequency side via a beat note
measurement \cite{FischHol00,FischDid00}:

\begin{equation}
2  f_n - f_{2n} = 2( n f_{\rm rep} + f_{\rm ceo}) - ( 2nf_{\rm
rep} + f_{\rm ceo}) = f_{\rm ceo}\, .
\end{equation}

\index{Frequency~comb!photonic crystal fiber}

\begin{figure}[t!]
\begin{center}
\includegraphics[width=0.95\textwidth]{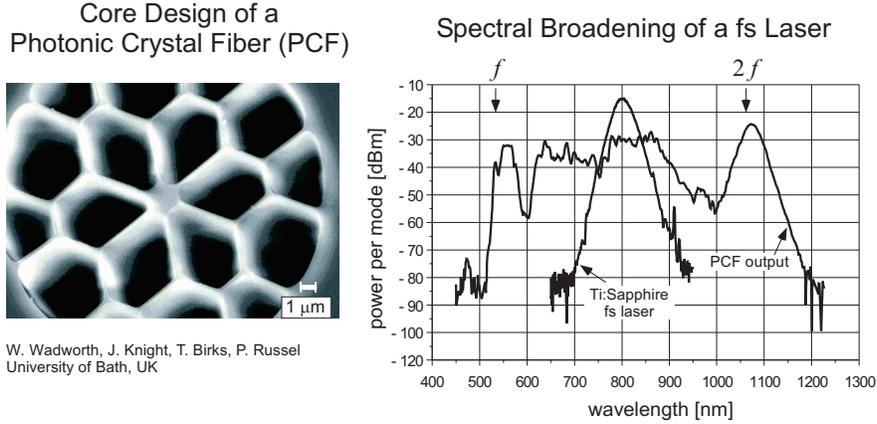}
\caption{  Core design of a photonic crystal fiber (PCF) and
spectral broadening of a fs laser. The PCF was seeded with 20 fs
pulses and the average output power of the PCF was 180 mW. }
\label{fischer6}
\end{center}
\end{figure}

If the spectrum does not cover an entire octave, one can
alternatively compare $3.5f_{8n}$ and $4f_{7n}$ to get
$\frac{1}{2}f_{\rm ceo}$ \cite{FischRei00a,FischRei99} or
$3f_{2n}$ with $2f_{3n}$ to obtain $f_{\rm ceo}$
\cite{FischRam02,FischMor01,FischFortier}. The broad spectra
needed for this technique are either directly emitted by the fs
laser \cite{FischBar02,FischEll01} or can be obtained by external
broadening in a highly nonlinear medium such as a photonic crystal
fiber (PCF) \cite{FischKni96,FischRan00}. A PCF as pictured in
Fig.\ref{fischer6} can be designed to have zero group velocity
dispersion (GVD) at 800 nm, which is the central wavelength of
commonly used Ti:sapphire fs lasers. Due to the vanishing GVD the
pulse spreading within the PCF is lower than in usual single mode
fibers. The resulting high peak intensity leads to self phase
modulation and therefore efficient broadening of the initial
frequency comb.

\begin{figure}[b!]
\begin{center}
\includegraphics[width=0.95\textwidth]{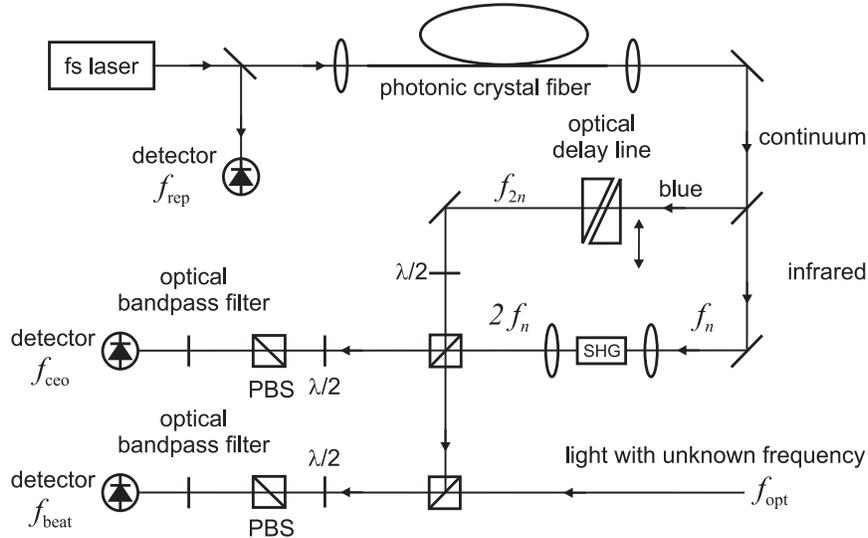}
\caption{ Experimental setup for detecting $f_{\rm rep}$, $f_{\rm
ceo}$ and $f_{\rm beat}$. An optical delay line is inserted into
the ``blue" arm of the nonlinear interferometer to match the
optical path lengths. PBS denotes a polarizing beam splitter.}
\label{fischer7}
\end{center}
\end{figure}

If $f_{\rm rep}$ and $f_{\rm ceo}$ are stabilized by phase
coherently linking them to a RF reference, the accuracy of the RF
reference is in one step transferred to all  cw modes of the
octave spanning optical frequency comb. Using state-of-the-art Cs
fountain clocks, which already reach accuracies of $10^{-15}$
\cite{FischMar03}, the frequency of an unknown light field can in
principle be measured with the same level of accuracy. The fs
frequency comb technique was tested to be accurate at the
$<10^{-16}$ level by comparing two independent systems
\cite{FischHol00,FischDid02}. To determine an optical frequency
$f_{\rm opt}$ of the unknown light field one  needs to measure the
frequency $f_{\rm beat}$ of the beat note between the unknown
light field and the neighboring mode $f_n$ of the frequency comb.
The unknown frequency $f_{\rm opt}$ can then be written as
\begin{equation}
f_{\rm opt} = f_n + f_{\rm beat} = n f_{\rm rep} + f_{\rm ceo} +
f_{\rm beat}\ .
\end{equation}
The mode number $n$ may be determined by a coarse measurement of
$f_{\rm opt}$ with a commercial wavemeter. Using the fs frequency
comb technique optical frequency measurements have been carried
out on atoms and ions, demonstrating accuracies of up to
$10^{-14}$ \cite{FischNie00a,FischUde01,FischEic03,FischSte02}. An
experimental setup for detecting $f_{\rm ceo}$, $f_{\rm rep}$ and
$f_{\rm beat}$ using an octave broad frequency comb is shown in
Fig.\ref{fischer7}.

\index{Atomic fountain}
\index{Absolute~frequency~measurements!of~optical~transitions}

Another application of fs frequency combs is the determination of
optical frequency ratios. As a frequency is dimensionless, no RF
reference based on Cs is needed and one can take advantage of the
high stability and accuracy of optical frequency standards, which
should lead to an increased sensitivity to the drift of
fundamental constants \cite{FischSte02}. Due to the invention of
photonic crystal fibers the complexity of the frequency
measurement in 2003 has been considerably reduced as compared to
the 1999 experiment, where a fs laser was already in use. The
experimental setup used in 2003 to measure the frequency of the
hydrogen spectroscopy dye laser was equivalent to that shown in
Fig.\ref{fischer7} and employed a fs Ti:sapphire ring laser
(GigaOptics, model GigaJet) with 800 MHz repetition rate. The
spectrum of the fs laser was externally broadened with the help of
a PCF to more than one octave including light from 946 nm to 473
nm. The detection of the repetition rate $f_{\rm rep}$ was placed
in front of the microstructured fiber to not be  affected by
amplitude noise caused by imperfect fiber coupling. $f_{\rm rep}$
was phase locked to a 800 MHz signal which was directly derived
from the transportable Cs fountain clock FOM. For both the 1999
and 2003 measurements, the transportable Cs fountain clock FOM has
been installed at MPQ. Its instability is $1.8\times
10^{-13}\tau^{-1/2}$ and its accuracy has been evaluated to be
$8\times 10^{-16}$ \cite{FischAbg03} at BNM-SYRTE. During the
experiments in Garching, only a verification at the level of
$10^{-15}$ has been performed. Consequently we attribute a
conservative FOM accuracy of $2\times10^{-15}$ for these
measurements.

\begin{figure}[t!]
\begin{center}
\includegraphics[width=0.95\textwidth]{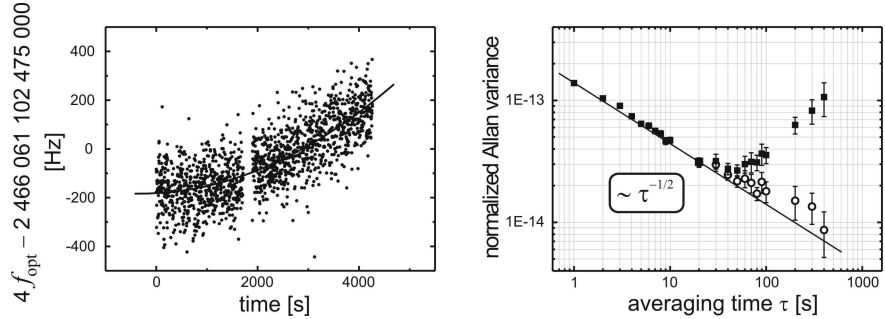}
\caption{ (left) beat frequency of the 486 nm dye laser relative
to the stabilized fs frequency comb. The solid line is a parabolic
fit to the data. (right) normalized Allan variance vs. averaging
time computed from a time series of 1 second counter readings with
a considerable dead time. The straight line indicates the
$\tau^{-1/2}$ dependence, which is the signature of the Cs
fountain clock. The raw data analysis (squares) shows that the
stability  for averaging times longer than 20 s is limited by the
drift of the ULE reference cavity. Open circles represent data
corrected for the parabolic cavity drift.} \label{fischer8}
\end{center}
\end{figure}

 To check for possible cycle slipping, the phase locked frequencies
$f_{\rm ceo}$ and $f_{\rm rep}$ were additionally counted to
verify consistency. The 486 nm dye laser and the blue part of the
frequency comb were spatially overlapped, optically filtered
around 486 nm, and directed onto an avalanche photodiode to
measure the beat frequency with the neighboring mode of the
frequency comb. The detected beat note was filtered, amplified and
directed to three radio frequency counters (Hewlett Packard,
models 53131A and 53132A) utilizing different detection bandwidth
and power level. All counters were referenced to the Cs clock. To
check for errors in the counting process only data points were
accepted where all three counter readings were consistent with
each other. Additionally it was verified that the dye laser was
successfully locked to the ULE reference cavity during the
measurement time.

Fig.\ref{fischer8} shows a typical beat note measurement (left)
and the corresponding normalized Allan variance (right) of the dye
laser locked to the reference cavity relative to the fs frequency
comb which was locked to FOM. For longer averaging times, the plot
of the Allan variance is generated by juxtaposing 1-s counter
readouts. Whereas it is known that such a procedure can alter the
functional dependence of the Allan variance \cite{FischLesage},
white frequency noise, as produced by the Cs fountain, is immune
to this form of bias. The observed $\tau^{-1/2}$ dependence
coincides with the independently measured fountain clock
instability for averaging times shorter than $\approx 10$~s. The
short term stability of the laser system is better than the
stability of the fountain clock. However, the long term stability
is limited by the drift of the ULE reference cavity.

To compensate for the slow ULE cavity drift we fit a second-order
polynomial to the measured beat note before averaging which
significantly reduces the Allan variance for longer averaging
times. To accurately determine the frequency of the dye laser, we
first average the frequency of the ULE cavity with a polynomial
such as the one shown on the left side of Fig.\ref{fischer7} with
the consistent counter readout. Then we use this polynomial and
the recorded  AOM readings for each data point, that determine the
cavity-laser detuning, to derive a highly stable value for the
laser frequency. For  the given stability of the Cs fountain clock
and the cavity, the optimum record length is around 500 s. For
longer averaging times the Cs fountain is more stable than the
drift-corrected ULE cavity.

\begin{figure}[t!]
\begin{center}
\includegraphics[width=0.7\textwidth]{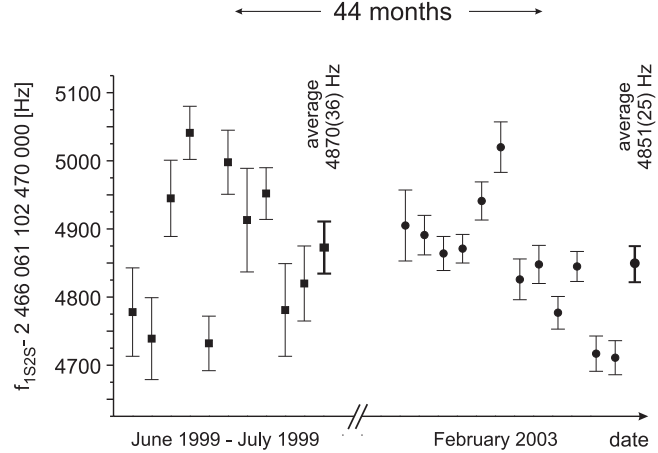}
\caption{ Experimental results and averages for the 1999 and 2003
measurements of the $(1S,\, F=1,\, m_F=\pm1\rightarrow 2S,\,
F'=1,\, m'_F=\pm1)$  transition frequency in atomic hydrogen.}
\label{fischer9}
\end{center}
\end{figure}

\index{Absolute~frequency~measurements!in~the~hydrogen~atom}
 We have measured
the $1S$--$2S$ transition in atomic hydrogen during 10 days in
1999 and during 12 days in 2003. Both data sets have been analyzed
using the same theoretical line shape model and are therefore
comparable. In Fig.\ref{fischer9}, the results of the
extrapolation to zero excitation light intensity and the
respective statistical error bars for each day are presented.
Since 1999, the statistical uncertainty for each day of
measurement was significantly reduced due to the narrower laser
linewidth and better signal-to-noise ratio, but the scatter of the
day averages did not reduce accordingly. We have tested several
possible reasons for this
\begin{table}[t!]
\begin{center}
\begin{tabular}{ l@{\hspace{2ex}\ }c@{\hspace{2ex}\ }c @{\hspace{2ex}\ }c@{\hspace{2ex}\ }c}
 \hline \rule{-5pt}{3ex}
Contribution&    $\nu_{H,1999}$ &$\sigma_{H,1999}$ &$\nu_{H,2003}$ & $\sigma_{H,2003}$ \\
     & [Hz]&  [Hz] & [Hz]& [Hz] \\
   \hline
Extrapolated value $-\ 2\,466\,061\,102\,474$ kHz&870&36&851&25\\
Background gas pressure shift &10&10&0&2\\
Intra-beam pressure shift &0&10&0&10\\
Lineshape model  &0&20&0&20\\
DC Stark shift &0&5&0&5\\
Blackbody radiation  &0&1&0&1\\
Standing wave effects &0&10&0&1\\
Intensity zero uncertainty  &0&1&0&0\\
Fountain clock uncertainty &0&5&0&5\\
%
 \hline
{Total $-\ 2\,466\,061\,102\,474$ kHz}&{ 880}&{45}&{ 851}&{ 34}\\
\hline
 \end{tabular}
  \caption{ { Results of the $(1S,\, F=1,\, m_F=\pm1\rightarrow 2S,\, F'=1,\, m'_F=\pm1)$
   transition frequency measurement ($\nu_{H,1999}$, $\nu_{H,2003}$) and uncertainty budgets ($\sigma_{H,1999}$, $\sigma_{H,2003}$) for the 1999 and 2003
   measurements correspondingly.}}
   \label{fischtab2}
   \end{center}
\end{table}
%
additional scatter including an intra-beam pressure shift, a
background gas pressure shift, Stark effects due to the RF gas
discharge, and DC Stark shift and have been able to exclude all
these effects at least on a conservative level of 10--20~Hz. A
possible origin of the observed  scatter can be due to a residual
first order Doppler effect arising from a violation of the axial
symmetry of the enhancement cavity mode and the hydrogen atomic
beam. The scattering of the excitation light on intra-cavity
diaphragms can also cause slight changes of the field distribution
and the corresponding first order Doppler effect. However, it
should average to zero over multiple adjustments of the hydrogen
spectrometer because the shifts can have both signs. As the
scatter is the same for both the measurement sets, we believe them
to be equivalent. The main statistical and systematic
uncertainties of these measurements are collected in Table
\ref{fischtab2}. The averaging of the 1999 and 2003 daily data points
was performed without weighting them.{\footnote {The result of
$2\,466\,061\,102\,474\,870$~Hz
 was inadvertently described in  \cite{FischNie00a} as ``the
weighted mean value'' but was calculated without consideration of
the daily statistical uncertainties.}} For  both measurements the
dominating resulting uncertainty arises from the day-to-day
scatter, while the pure statistical uncertainty for each day  is
significantly smaller. In fact, weighting of the day data only
slightly influences the results (on the level of $\sigma/2$).

Comparing both measurements we deduce a difference of
$\nu_{H,2003}-\nu_{H,1999}$ equal to $(-29\pm 57)$~Hz within 44
months. This corresponds to a relative drift of $\nu_{\rm {H}}$
against the $^{133}$Cs ground state hyperfine splitting of
$\partial_t (\ln(\nu_{\rm Cs}/\nu{\rm_{H}}))=(3.2\pm
6.3)\times 10^{-15}$ per year.

\section{Determination of drift rates}

Despite the high sensitivities (less than $10^{-14}$~yr$^{-1}$),
the accuracy of transition frequency drift measurements are rather
low (uncertainty is typically over $100$\%), so that only the
first order expansion in terms of the constants involved in the
evaluation is sufficient. The frequency of any optical transition
can be written as
\begin{equation} \label{Fischfour}
\nu=const\,   Ry \,  F_{\rm rel}(\alpha),
\end{equation}
where  $Ry$ is the Rydberg frequency expressed in Hertz and $F_{\rm
rel}(\alpha)$ takes into account relativistic and many-body
effects. The Rydberg energy cancels in atomic frequency
comparisons. Therefore the dependence of $Ry$ on $\alpha$
($Ry\sim\alpha^2$) and other fundamental constants contained in
$Ry$ is irrelevant{\footnote {The expansion of $\nu$ in terms of
small changes of $\alpha$ as given in \cite{FischDzu99} are said
to be derived assuming the constancy of the Rydberg frequency.
However, no such restriction on the unit of frequency is necessary
here, as any choosen unit will cancel out in the final result
since only frequency ratios are used.}}. The relativistic
correction $F_{\rm rel}$ depends on the transition in the system
considered and embodies additional dependence on $\alpha$, while
$const$ is a numerical factor and is independent of any
fundamental constants.

\index{Absolute~frequency~measurements!in~the~mercury~ion}

The frequency
\begin{equation}
\nu_{\rm {Hg}}=1\ 064\ 721\ 609 \ 899 \ 143.7 (10) \ {\rm Hz}
\end{equation}
of the $5d^{10}6s \ ^2S_{1/2} (F=0)\rightarrow 5d^9 6s^2\
^2D_{5/2} (F'=2, m'_F=0)$ electric quadrupole transition in
$^{199}$Hg$^+$ was precisely measured at NIST between the years
2000 and 2002 \cite{FischBiz03}. Numerical calculations including
relativistic and many-body effects for the dependence of $F_{\rm
rel, {\rm Hg}}(\alpha)$ for $\nu_{\rm {Hg}}$ on the fine structure
constant $\alpha$ yield \cite{FischDzu99}
\begin{equation}
\alpha \frac{\partial}{\partial \alpha}\ln F_{\rm rel, {\rm
Hg}}(\alpha) \approx -3.2\,.
\end{equation}

In the light hydrogen atom, the relativistic correction for
$\nu_{\rm {H}}$ nearly vanishes ($F_{\rm rel, {\rm H}}(\alpha) \approx {\rm const.}$):
\begin{equation}
\alpha \frac{\partial}{\partial \alpha}\ln F_{\rm rel, {\rm
H}}(\alpha) \approx 0
\end{equation}
or
\begin{equation}
\nu_{\rm {H}}\sim Ry\, .
\end{equation}

The frequency of hyperfine transitions have a different functional
dependence on $\alpha$. For the ground state hyperfine transition
in $^{133}$Cs we have
\begin{equation} \nu_{\rm {Cs}} = const'\  Ry \
\alpha^2\  \frac{\mu_{\rm {Cs}}}{\mu_{B}} \  F_{\rm rel, {\rm
Cs}}(\alpha) \label{Fischten}
\end{equation}
with a relativistic correction $F_{\rm rel, {\rm Cs}}(\alpha)$ of
\cite{FischDzu99}
\begin{equation}
\alpha \frac{\partial}{\partial \alpha}\ln F_{\rm rel, Cs}(\alpha)
\approx +0.8\,.
\end{equation}

Combining these equations, we find that the comparison of the
clock transition in Hg against a primary frequency standard tests
the following fractional time variation \cite{FischBiz03}:
\begin{eqnarray}
 \frac{\partial}{\partial t} \ln\frac{\nu_{\rm {Cs}}}{\nu_{\rm {Hg}}}
 & = & \frac{\partial}{\partial t} \ln \left( \frac{\alpha^2 \frac{\mu_{\rm {Cs}}}{{\mu_{B}}} F_{\rm rel, {\rm Cs}}(\alpha)}{F_{\rm rel, {\rm Hg}}(\alpha)} \right)
   =  2 \frac{\partial \ln \alpha}{\partial t}  +  \frac{\partial}{\partial t} \ln\frac{\mu_{\rm {Cs}}}{\mu_{B}} +
   (0.8+3.2)  \frac{\partial \ln \alpha}{\partial t} \nonumber \\
 & = & 6 \frac{\partial}{\partial t} \ln \alpha  +  \frac{\partial}{\partial t} \ln\frac{\mu_{\rm {Cs}}}{\mu_{B}}
   = (0.2 \pm 7) \times 10^{-15}\ \textrm{yr}^{-1}
\end{eqnarray}

Likewise we derive for the fractional variation of $\nu_{\rm
{Cs}}/\nu_{\rm {H}}$ from the hydrogen $1S$--$2S$ experiment
\textrm{[this work]}:
\begin{equation}
\frac{\partial}{\partial t} \ln\frac{\nu_{\rm {Cs}}}{\nu_{\rm
{H}}} = 2.8 \frac{\partial}{\partial t} \ln \alpha +
\frac{\partial}{\partial t} \ln\frac{\mu_{\rm {Cs}}}{\mu_{B}} =
(3.2 \pm 6.3) \times 10^{-15}\ \textrm{yr}^{-1} \ \
\end{equation}

With $x=\partial_t \ln\alpha$ and $y=\partial_t \ln(\mu_{\rm {Cs}}/{\mu_{B}})$ we can write the experimental
results as
\begin{eqnarray}
\label{fischeqarr1} 6x + y & = & (0.2 \pm 7) \times 10^{-15}\
\textrm{yr}^{-1} \qquad \textrm{(Hg}^+\textrm{),} \\
2.8x +y & = & (3.2 \pm 6.3) \times 10^{-15} \ \textrm{yr}^{-1}
\qquad \textrm{(H).}\label{fischeqarr2}
\end{eqnarray}

\begin{figure}[t]
\begin{center}
\includegraphics[width=0.8\textwidth]{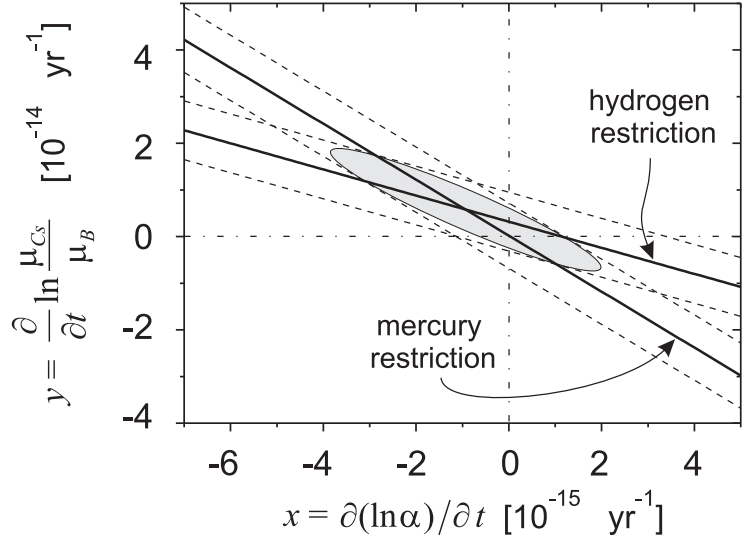}
\caption{ Drifts of the $^2S_{1/2}(F=0)\rightarrow {^2}D_{5/2}
(F'=2, m'_F=0)$ transition in $^{199}$Hg$^+$ and of the $1S(F=1,
m_F=\pm 1)\rightarrow2S(F'=1,m'_F=\pm 1)$ transition in H against
the frequency of the ground-state hyperfine transition in
$^{133}$Cs.  Dashed lines represent 1 $\sigma$ experimental
restrictions from the mean measured values. The elliptical region
defined by $R(\Delta x, \Delta y)=1$ gives the standard deviation
for $x$ and $y$ when projected on corresponding axis by
integration over the other.} \label{fischer0}
\end{center}
\end{figure}

\index{Model-independent evaluation}

These equations are easily solved, yielding the mean expectation
values $\langle y \rangle$ and $\langle x \rangle$ without any
assumptions of possible correlations between the drifts. In
Fig.\ref{fischer0}, both equations and the graphical solution are
shown. Obviously, testing the stability of $\alpha$ by monitoring
only one transition frequency during a time period would require
additional assumptions of the drift of other fundamental
constants.

The uncertainties can be calculated by making two assumptions: (i)
the experimental data are Gaussian distributed and (ii) the
mercury (\ref{fischeqarr1}) and the hydrogen (\ref{fischeqarr2})
measurements are statistically independent. In this case normal
gaussian error propagation allows the calculation of the variances
$\langle y^2 \rangle-\langle y \rangle^2$ and $\langle x^2
\rangle-\langle x \rangle^2$ even when the drift rates $x$ and $y$
are correlated \cite{FischCal02}. This is because the covariance
term $\langle xy\rangle$ does not appear when
(\ref{fischeqarr1},\ref{fischeqarr2}) are resolved for $x$
and $y$.

For a graphical representation it is possible to calculate the
two-dimensional probability density of $x$ and $y$ to be the true
values:
\begin{equation}
P(x,y)=\frac{1}{2\pi\sqrt{\sigma_{\rm H}\, \sigma_{\rm
Hg}}}\exp[{-{R(\Delta x, \Delta y)}/{2}}],
\end{equation}
where $\Delta x$ and $\Delta y$ are the distances along the
corresponding axes from the crossing point of the solid lines
(Fig.\ref{fischer0}) i.e. the solution of
(\ref{fischeqarr1},\ref{fischeqarr2}). The experimental
uncertainties are {$\sigma_{\rm H}=6.3\times 10^{-15}$~yr$^{-1}$}
and $\sigma_{\rm Hg}=7\times 10^{-15}$~yr$^{-1}$ taken from
(\ref{fischeqarr1}) and (\ref{fischeqarr2}), and the exponent
function is given by:
\begin{equation}
{R(\Delta x, \Delta y)}=(\Delta y+6\Delta x)^2/\sigma_{\rm
Hg}^2+(\Delta y+2.8\Delta x)^2/\sigma_{\rm H}^2\, .
\end{equation}
We deduce the uncertainties for $x$ and $y$  as projections of the
ellipse defined by $R(\Delta x, \Delta y)=1$
 on the corresponding axes (Fig.\ref{fischer0}) by integration
 over the other dimension. For only two independent measurements
 this method is equivalent to performing simple Gaussian error
 propagation of uncertainties when resolving
 (\ref{fischeqarr1},\ref{fischeqarr2}). However, the
 projection method can be generalized to more than two
 measurements, i.e. more than two equations for the two unknowns
 $x$ and $y$ (see contribution by E.~Peik in this volume). The
 integration in both directions can be performed analytically to
 derive the uncertainties of $x$ and $y$.
Our evaluation is model-independent in the sense that we neither
assume $x$ and $y$ to be uncorrelated nor that they are correlated
in any way.

The relative drift of the fine structure
constant $\alpha$ between July 2000 and the end of 2003 is
\begin{equation}
x=\frac{\partial}{\partial t}{\ln \alpha} = (-0.9 \pm 2.9) \times
10^{-15} \ {\rm yr}^{-1}. \label{Fischalph}
\end{equation}
For the limit on the relative drift of $\mu_{\rm {Cs}}/\mu_{B}$,
we find
\begin{equation}\label{Fischmu}
y=\frac{\partial}{\partial t} \ln \frac{\mu_{\rm {Cs}}}{\mu_{B}}=
(0.6 \pm 1.3) \times 10^{-14} \ {\rm yr}^{-1}.
\end{equation}

The given 1 $\sigma$ uncertainties for $x$ and $y$
 incorporate both the
statistical and systematic uncertainties of the hydrogen and the
mercury measurements.  Both limits (\ref{Fischalph}) and
(\ref{Fischmu}) are consistent with zero.

\index{Rb--Cs clock comparison}

These results allow us to deduce a restriction for the relative
drift of the ratio of the nuclear magnetic moments in $^{87}$Rb
and $^{133}$Cs. From 1998 to 2003, the drift of the ratio of the
ground state hyperfine frequencies in $^{87}$Rb and $^{133}$Cs has
been measured to be \cite{FischMar03}
\begin{eqnarray}
\frac{\partial}{\partial t} \ln\frac{\nu_{\rm {Rb}}}{\nu_{\rm
{Cs}}}= (0.2 \pm 7.0) \times 10^{-16}\ \textrm{
yr}^{-1}.\label{Fischclock}
\end{eqnarray}
Substituting the corresponding dependencies $F_{\rm rel}(\alpha)$
for these transitions \cite{FischMar03,FischDzu99}, we can write
\begin{equation}
\frac{\partial}{\partial t} \ln\frac{\nu_{\rm {Rb}}}{\nu_{\rm
{Cs}}}=\frac{\partial}{\partial t}\left( \ln\frac{\mu_{\rm
{Rb}}}{\mu_{\rm {Cs}}}-0.53\ln\alpha\right). \label{Fischmagmom}
\end{equation}
Combining (\ref{Fischalph}), (\ref{Fischclock}), and
(\ref{Fischmagmom}) we deduce a restriction for the relative drift
of the nuclear magnetic moments in $^{87}$Rb and $^{133}$Cs:
\begin{equation}\label{Fischratio}
\frac{\partial}{\partial t} \ln\frac{\mu_{\rm {Rb}}}{\mu_{\rm
{Cs}}}=(-0.5\pm1.7)\times10^{-15}\ \textrm{yr}^{-1}.
\end{equation}
where the  same procedure as in Fig.\ref{fischer0} was used with a
diagram covering $x$ and $z=\partial_t\ln(\mu_{\rm
Rb}/\mu_{\rm Cs})$.

\index{Variation~of~fundamental~constants!gp@$g_p$}

 The values of
the nuclear moments are determined by the strong and the
electromagnetic interaction. If the former is constant, the time
changing nuclear moments point toward a variation of the strong
coupling constant. Unfortunately, there is no simple scaling law
such as (\ref{Fischfour}) or (\ref{Fischten}) known for the
nuclear moments. However, they can be approximated with the
Schmidt model \cite{FischKar00}. For $^{87}$Rb and $^{133}$Cs
atoms the Schmidt nuclear magnetic moments $\mu^{s}$ depend only
on the proton gyromagnetic ratio $g_p$. Using this model, one can
get an approximate relation
\begin{equation}
\frac{\partial}{\partial \ln g_p} \ln\frac{\mu_{\rm
{Rb}}^{s}}{\mu_{\rm {Cs}}^{s}}\simeq 2,
\end{equation}
which, in combination with (\ref{Fischratio}), yields a stringent
upper bound for the drift of the proton gyromagnetic factor $g_p$:
\begin{equation}
\frac{\partial}{\partial t} \ln g_p = (-0.2\pm0.8)\times10^{-15}\
\textrm{yr}^{-1}.
\end{equation}

\begin{table}[t!]
\begin{center}
\begin{tabular}{ c@{\hspace{1ex}\ }c@{\hspace{1ex}\ }c@{\hspace{1ex}\ }c@{\hspace{1ex}\ }c}
 \hline \rule{-5pt}{3ex}
  { Method, reference} &   $t_2-t_1$&$[\alpha(t_1)-\alpha(t_2)]/\alpha$   & { Model assumptions} \\
\hline\rule{-5pt}{3ex}
Geological   & 2~Gyr&$(-0.36\pm1.44)\times 10^{-8}$&fission conditions,\\
(Oklo reactor) \cite{FischFuj00}&&& ${\dot \alpha_{S}}={\dot\alpha_{W}}=0$\\
\rule{-5pt}{3ex}
Astrophysical   & 5--11~Gyr&$(-0.54\pm0.12)\times10^{-5}$&astrophysical \\
(absorption spectra) \cite{FischMur03}      & && models\\
\rule{-5pt}{3ex}
Astrophysical   & 9.7~Gyr&$(-0.06\pm0.06)\times10^{-5}$&astrophysical \\
(absorption spectra) \cite{FischChand}      & && models\\
\rule{-5pt}{3ex}
Astrophysical   & 8~Gyr&$(0.1\pm1.7)\times10^{-5}$&astrophysical \\
(absorption spectra) \cite{FischQuast}      & && models\\
\rule{-5pt}{3ex}
Laboratory (Rb--Cs  & 4~yr&$(0.2\pm5.2)\times10^{-15}$&${\dot \alpha_{S}}={\dot\alpha_{W}}=0$\\
clocks comparison) \cite{FischMar03}      & && \\
\rule{-5pt}{3ex}
Laboratory & 3~yr&$(-0.1\pm3.5)\times10^{-15}$&${\dot \alpha_{S}}={\dot\alpha_{W}}=0$\\
(Hg$^{+}$ transition  frequency &&&\\
   measurement) \cite{FischBiz03}      & && \\
\rule{-5pt}{3ex}
Laboratory  & 3.6~yr&$(-4.1\pm 8.2)\times10^{-15}$&${\dot \alpha_{S}}={\dot\alpha_{W}}=0$\\
(H transition frequency&&&\\
    measurement) [this work]       & && \\

\hline\rule{-5pt}{3ex}
\rule{-5pt}{3ex}
Combination of \cite{FischBiz03}  & 3.5~yr&$(3.2\pm10.2)\times10^{-15}$&LLI, LPI,\\
   and this work      & &&linear drifts \\
\hline
 \end{tabular}
  \caption{ {
Some of the precise recent measurements testing the relative changes
of the fine-structure constant $\alpha$ over a time interval
$(t_2-t_1)$ where $t_2$ is the present time and $t_1$ corresponds
to the past. The drift  can be calculated as $\partial/\partial
t(\ln\alpha)\simeq[\alpha(t_2)-\alpha(t_1)]\ \alpha^{-1}\
(t_2-t_1)^{-1}$. Combining the results of absolute frequency
measurements of the optical transitions in Hg$^{+}$ and H yields a
 restriction for the drift of $\alpha$ without assumptions
  of conceivable correlations between the constants.}}
   \label{fischtab1}
   \end{center}
\end{table}

\section{Conclusion}

In conclusion, we have determined limits for the drift of
$\alpha$, $\mu_{\rm {Cs}}/\mu_{B}$ and $\mu_{\rm {Rb}}/\mu_{\rm
{Cs}}$ from laboratory experiments without any assumptions of
their conceivable correlations. All these limits are consistent
with zero drift. Table \ref{fischtab1} represents some of the most
accurate recent measurements of drifts of the fine structure
constant $\alpha$ in different  epochs. From all these data only
the investigations of quasar absorption spectra measured with the
Keck/HIRES spectrograph show a significant deviation between the
values of $\alpha$ today and 10 Gyrs ago \cite{FischMur03}.
Considering the Oklo data as well as results of modern
astrophysical and laboratory measurements one can suppose that the
drift, if existent at all, is not linear and that $\alpha$ has
reached an asymptotic value or is in the extremum of an
oscillation or is simply too small to be detected yet. To make a
definite conclusion additional independent astrophysical data as
well as a further increase of the accuracy of laboratory methods
are required.

\section{Acknowledgements}

We thank S.G.~Karshenboim  for the fruitful discussion of this
work. N.~Kolachevsky acknowledges  support from Alexander von
Humboldt Stiftung. The work was partly supported by the Deutsche
Forschungsgemeinschaft (grant No. 436RUS113/769/0-1) and RFBR. The
development of the FOM fountain was supported by Centre National
d'\'{e}tudes spatiales and Bureau National de M\'{e}tro\-logie.

\end{document}